\documentclass{PoS}

\title{Extragalactic cosmic ray self-confinement around sources}

\ShortTitle{Cosmic ray self-confinement}

\author{\speaker{Pasquale Blasi}\\
        INAF/Osservatorio Astrofisico di Arcetri, largo E. Fermi 5, Firenze, Italy\\
       Gran Sasso Science Institute (INFN), Viale F. Crispi 7, L'Aquila, Italy\\
        E-mail: \email{blasi@arcetri.astro.it}}
        
\author{{Elena Amato}\\
        INAF/Osservatorio Astrofisico di Arcetri, largo E. Fermi 5, Firenze, Italy\\
        E-mail: \email{amato@arcetri.astro.it}}

\author{{Marta D'Angelo}\\
        Gran Sasso Science Institute (INFN), Viale F. Crispi 7, L'Aquila, Italy\\
        E-mail: \email{marta.dangelo@gssi.infn.it}}

\abstract{Most models of the origin of ultra high energy cosmic rays rely on the existence of luminous extragalactic sources. Cosmic rays escaping the galaxy where the source is located produce a sufficiently large electric current to justify the investigation of plasma instabilities induced by such current. Most interesting is the excitation of modes that lead to production of magnetic perturbations that may scatter particles thereby hindering their escape, or at least changing the propagation mode of escaping cosmic rays. We argue that self-generation of waves may force cosmic rays to be confined in the source proximity for energies $E\lesssim 10^{7} L_{44}^{2/3}$ GeV for low background magnetic fields ($B_{0}\ll nG$). For larger values of $B_{0}$, cosmic rays are confined close to their sources for energies $E\lesssim 2\times 10^{8} \lambda_{10} L_{44}^{1/4} B_{-10}^{1/2}$ GeV, where $B_{-10}$ is the field in units of $0.1$ nG, $\lambda_{10}$ is its coherence length in units of 10 Mpc and $L_{44}$ is the source luminosity in units of $10^{44}$ erg/s. }

\FullConference{The 34th International Cosmic Ray Conference,\\
		30 July- 6 August, 2015\\
		The Hague, The Netherlands}

\newcommand{\be}{\begin{equation}}
\newcommand{\ee}{\end{equation}}

\begin{document}

\section{Introduction}

The onset of extragalactic cosmic rays and the end of the Galactic cosmic ray (CR) component are subjects of major investigation at this time. However this transition region remains poorly understood \cite{Rendus} and the conclusions we draw depend rather sensibly on the mass composition and on the spectral shape of extragalactic CRs. Of particular importance is the fact that in most models of the transition from Galactic to extragalactic CRs one is forced to assume that the flux of the extragalactic component is suppressed at sufficiently low energies. Some constraints on the sources of ultra high energy CRs (UHECRs) can be obtained as follows \cite{waxman}: for a source with size $R$ to be able to accelerate CRs to energy $E$, the Larmor radius $E/ZeB$ must be smaller than the size $R$ (here $Ze$ is the charge of CRs and $B$ is the magnetic field). This implies that $B>E/ZeR$. Since the source must be able to guarantee at least a magnetic energy flux $v B^{2}/4\pi$, where $v$ is the velocity of the accelerator, one easily gets a lower limit on the source luminosity: $L\gtrsim 3.2 \times 10^{45} Z^{-2} \left(\frac{E}{10^{20} eV}\right)^{2} \beta ~ \rm erg/s$ where $\beta=v/c$ is the dimensionless velocity in units of the speed of light. The required luminosity becomes even higher in the case of relativistic sources \cite{waxman}. The general conclusion is that the sources of UHECRs have to be very luminous. 

Being interested in investigating the escape of CRs from their sources, here we will not consider the problem of acceleration any longer and we will concentrate on the propagation outside the host galaxy inside which the actual source is located. On the other hand, for the sake of brevity we will refer to such galaxy as the  {\it source}.

Let us assume, for simplicity, that CRs leave their host galaxy with an injection spectrum $q(E)\propto E^{-2}$ up to some maximum energy $E_{\rm max}$. The differential number density of CRs streaming out of such sources can then be written as:
$$
n_{\rm CR}(E,r) = \frac{q(E)}{4\pi r^{2} c}=\frac{L_{\rm CR}}{\Lambda} \frac{E^{-2}}{4 \pi r^2c}\approx 
$$
\be
\approx 1.7\times 10^{-14} L_{44}\ E_{\rm GeV}^{-2}\ r_{\rm Mpc}^{-2}\  \rm cm^{-3} \rm GeV^{-1}\ ,  
\label{eq:nbal}
\ee
where we have adopted $\Lambda=\ln(E_{\rm max}/E_{\rm min})\approx 25$ and $L_{\rm CR}=10^{44} L_{44}$ erg/s, energies are in GeV and distances in Mpc. The source is assumed to be located in an average position of the intergalactic medium (IGM) with a density of baryonic gas is $n_{b}=\Omega_{b}\rho_{cr}/m_{p}=2.5\times 10^{-7}~\left(\frac{\Omega_{b}h^{2}}{0.022}\right)\rm cm^{-3}$. We also assume that there is a cosmological magnetic field with a strength  $B_{0}=10^{-13}B_{-13}$ G and a correlation scale of the order of $\lambda_{B}\sim 10$ Mpc, so that on scales smaller than $\lambda_{B}$, the field can be considered as oriented along a given $\hat z$ direction. Hence the Alfv\'en speed is $v_{A}=B_{0}/\sqrt{4\pi \Omega_{b}\rho_{cr}} \approx 44 ~B_{-13}\rm cm~s^{-1}$.

Assuming that CRs escaping the host galaxy are all protons (positively charged particles), one can easily see how this situation leads to the formation of an electric current directed outwards. In order to avoid charge accumulation anywhere in the Universe, one is forced to require a return current that exactly compensates the CR current. This situation, similar to that the takes place near supernova remnants, is known to be able to potentially lead to instabilities. Investigating such instabilities and their implications is the main purpose of this paper. 

\section{Calculations}

The current of CRs streaming away from their source into the IGM can be written as 
\be
J_{\rm CR}=e n_{\rm CR}(>E) c=\frac{e L_{\rm CR}E^{-1}}{4\pi \Lambda r^{2}},
\label{eq:current}
\ee
where $E$ is the minimum energy of the particles constituting the current. The validity of this expression is restricted to the case of vanishing background magnetic field, or a case in which the Larmor radius of the particles is $r_L\gg\lambda_B$, but we shall see that the above estimate for the current density turns out to hold also in the diffusive regime.

A current propagating in a plasma can give rise to instabilities of different types. Granted that the current carrying particles are well magnetised ($v_A>J_{\rm CR}/(e n_b)$, which from Eq.~\ref{eq:current} is seen to imply $B_0>2 \times 10^{-13}\ L_{44} E_{\rm GeV}^{-1} r_{\rm Mpc}^{-2}$)~G, the fastest growing instability arises when the condition 
\be
J_{\rm CR}E>\frac{ce B_0^2}{4\pi}
\label{eq:cond1}
\ee
is satisfied. This condition, which is the standard one for the development of non-resonant modes of the streaming instability \cite{bell2004}, is easily seen to be equivalent to the requirement that the energy density of CRs be locally larger than the energy density in the form of pre-existing magnetic field, $B_{0}^{2}/4\pi$. For $q(E)\propto E^{-2}$, this requirement becomes independent of $E$ and, using Eq.~\ref{eq:current}, can be simply formulated as:
\be
r <  r_{\rm inst} =3.7\times 10^{4} \frac{L_{44}^{1/2}}{B_{-13}}~ \rm Mpc.
\label{eq:nonres}
\ee
When Eq.~\ref{eq:nonres} is satisfied, the fastest growing modes have a wavenumber $k_{\rm max}$ that reflects the equilibrium between magnetic tension and $J_{\rm CR}\times \delta B$ force on the plasma, namely $k_{\rm max}B_{0} = \frac{4\pi}{c}J_{\rm CR}$, and their growth rate is:
\be
\gamma_{\rm max} = k_{\rm max} v_{A}=\sqrt{\frac{4\pi}{n_{b} m_{p}}} \frac{J_{\rm CR}}{c}\ ,
\label{eq:growthrate}
\ee
independent of the initial value of the local magnetic field $B_{0}$. The scale of the fastest growing modes $k_{\rm max}^{-1}$ is much smaller than the Larmor radius of the particles dominating the current (this is entailed in the condition for growth, Eq.~\ref{eq:cond1}), therefore they have no direct influence on particle scattering. This conclusion is however changed by the non-linear evolution of the modes. As long as the instability develops on small scales, it cannot affect the current, hence one could treat the two as evolving separately. A fluid element will be subject to a force that is basically $\sim J_{\rm CR}\delta B/c$: its equation of motion is $\rho (dv/dt) \simeq \frac{1}{c} J_{\rm CR} \delta B$, with $\delta B(t)=\delta B_{0} \exp\left(\gamma_{\rm max} t\right)$. One can estimate the velocity of a fluid element as $v \sim (\delta B(t) J_{\rm CR})/(c \rho \gamma_{\rm max})$, which upon integration in time leads to an estimate of the mean fluid displacement as $\Delta x \sim (\delta B(t) J_{\rm CR})(c \rho \gamma_{\rm max}^{2})$. The saturation of the instability is achieved when the displacement equals the Larmor radius of particles in the current, as calculated in the amplified magnetic field, $E/e \delta B$: when this condition is fulfilled, scattering becomes efficient and the current is destroyed. This simple criterion returns the condition:
\be
\frac{\delta B^{2}}{4\pi} \approx \frac{J_{\rm CR} E}{c e}=n_{\rm CR}(>E) E.
\label{eq:equi}
\ee
Since $n_{\rm CR}(>E)\propto E^{-1}$ in the case considered here, the saturation value of the magnetic field is independent of the energy of particles in the current driving the instability. A somewhat lower value of the saturation can be inferred following the criteria discussed in \cite{riquelme}, where the non-linear increase of the wavelength of the fastest growing modes is followed. Using such a prescription our saturation magnetic field would be $\sim 10$ times smaller. Eq.~\ref{eq:equi} expresses the condition of equipartition between the CR energy density and the amplified magnetic pressure, a condition that is often assumed in the literature without justification, and that here arises as a result of the physics of magnetic amplification itself. 

The field strength, as a function of the distance $r$ can now be written as
\be
\delta B(r) = 3.7 \times 10^{-9} L_{44}^{1/2} r_{\rm Mpc}^{-1} ~\rm Gauss\ .
\label{eq:magnetic}
\ee
This rather strong magnetic field will develop over distances $r$ from the source that satisfy Eq.~\ref{eq:nonres} and under the additional condition that the growth is fast enough so as to reach saturation in a fraction of the age of the universe, $t_0$ (numerical simulations of the instability \cite{bell2004} show that saturation occurs when $\gamma_{max}\tau \sim 5-10$). This latter condition reads $\gamma_{\rm max} t_0\gtrsim 5$ and translates into:
\be
r<r_{\rm growth}=1.2 \times 10^{4}L_{44}^{1/2}E_{\rm GeV}^{-1/2}\ \rm Mpc\ .
\label{eq:growth}
\ee
If the conditions expressed by Eqs.~\ref{eq:nonres} and \ref{eq:growth} are fulfilled, then the magnetic field can be estimated as in Eq.~\ref{eq:magnetic} and since $\delta B\gg B_{0}$ and there is roughly equal power at all scales, it is reasonable to assume that particle propagation can be described as Bohm diffusion in the magnetic field $\delta B$, so that 
\be
D(E,r)=9\times 10^{24} E_{\rm GeV}\ r_{\rm Mpc}\ L_{44}^{-1/2} \rm cm^{2}~s^{-1}.
\ee
The initial assumption of ballistic propagation of CRs escaping a source leads to conclude that particles would produce enough turbulence to make their motion diffusive. The diffusion time over a distance $r$ from the source can be estimated as
$\tau_{d}(E,r) = r^{2}/D(E,r) \approx 3.3\times 10^{16} r_{\rm Mpc}\ E_{\rm GeV}^{-1}\ L_{44}^{1/2} \rm yr$
from which follows that particles can be confined within a distance $r$ from the source for a time exceeding the age of the Universe, if their energy satisfies the condition:
\be
E\lesssim E_{\rm conf}=2.6\times 10^{6}\ r_{\rm Mpc}\ L_{44}^{1/2}\ \rm GeV\ .
\label{eq:confine}
\ee

One might argue that this conclusion contradicts the assumptions of our problem: for instance the density of particles in the diffusive regime is no longer as given in Eq.~\ref{eq:nbal}. This is certainly true, but the current that is responsible for the excitation of the magnetic fluctuations remains the same, as can easily be demonstrated: for particles with energy $E>E_{\rm conf}$ in Eq.~\ref{eq:confine}, and assuming that energy losses are negligible, quasi-stationary diffusion can be described by the equation
\be
\frac{1}{r^{2}}\frac{\partial}{\partial r}\left[ r^{2} D(E,r) \frac{\partial n}{\partial r}\right] = \frac{q(E)}{4\pi r^{2}}\delta (r),
\ee
where $q(E)$ is the injection rate of particles with energy $E$ at $r=0$. This equation is easily solved to provide the density of CRs at distance $r$ from the source:
\be
n(E,r) \approx \frac{q}{8\pi r D(E,r)}.
\label{eq:ndiff}
\ee
Clearly, by definition of diffusive regime, the density of particles returned by Eq.~\ref{eq:ndiff} is larger than the density in the ballistic regime, Eq.~\ref{eq:nbal}. However, the current in the diffusive regime is 
\be
J_{\rm CR}^{\rm diff}= e E D(E,r) \frac{\partial n}{\partial r} = e\frac{q(>E)}{4\pi r^{2}},
\label{eq:diffcur}
\ee
which is exactly the same current that we used in the case of ballistic CR propagation (see Eq.~\ref{eq:current}). This is a very important and general result: the magnetic field in Eq.~\ref{eq:magnetic} is reached outside a CR source independent of the mode of propagation of CRs, since it is only determined by the current and not by the absolute value of the CR density. Clearly the particles that are confined within a distance $r$ around the source do not contribute to the CR current at larger distances. 

\section{Results and implications}

The energy below which CRs can be considered as confined in the source vicinity can be calculated by taking into account all the three conditions that need to be imposed to guarantee such a confinement, namely Eq.~\ref{eq:nonres} (existence of fastly growing modes), Eq.~\ref{eq:growth} (growth rate faster than the expansion of the universe) and Eq.~\ref{eq:confine} (confinement). The first condition yields a limit on the distance from the source that is easy to satisfy, unless the strength of the background magnetic field is increased by several orders of magnitude, in which case however other complications arise (see discussion below). 

The other two conditions lead to the constrain
\be
E_{\rm cut} \approx 10^{7}\ {\rm GeV}\ L_{44}^{2/3} .
\label{eq:Ecut}
\ee
These particles are confined within a distance from the source:
\be
r_{\rm conf} \approx 3.8\ {\rm Mpc}\ L_{44}^{1/6} .
\label{eq:Rconf}
\ee
Within such a distance the magnetic field is as given by Eq.~\ref{eq:magnetic} and larger than $\delta B(r_{\rm conf})\approx 9.6 \times 10^{-10} L_{44}^{1/3}$ G. It is noteworthy that both the size of the confinement region and the magnetic field depend weakly upon the CR luminosity of the source, respectively as $L_{44}^{1/6}$ and $L_{44}^{1/3}$. Hence we can conclude that magnetic fields at the level of $0.1-1$nG must be present in regions of a few Mpc around the sources of UHECRs. As a consequence, the spectrum of CRs leaving these sources and eventually reaching the Earth must have a low energy cutoff at an energy $E_{cut}$. This kind of cutoff has been often postulated in the literature in order to avoid some phenomenological complications that affect models for the origin of UHECRs. For instance, a low energy cutoff is required in the dip model \cite{dip1,dip2} to describe appropriately the transition from Galactic to extragalactic CRs. This feature is usually justified by invoking some sort of magnetic horizon in the case that propagation of UHECRs is diffusive in the lower energy part of the spectrum \cite{horizon}. A similar low energy suppression of the CR flux is required by models with a mixed composition \cite{mixed}. In the calculations illustrated above, the presence of nuclei is readily accounted for, provided the current is still produced by protons (assumed to be the most abundant specie). In this case, the value of $E_{\rm cut}$ is simply shifted to $Z$ times higher energy for a nucleus of charge $Z$. The model proposed here provides a mechanism of self-confinement of CRs close to their sources that does not need artificial assumptions on diffusive propagation.

As discussed above, the condition that guarantees the existence of non-resonant modes (Eq.~\ref{eq:nonres}) is easily satisfied, unless the background magnetic field reaches $B_0\approx 9.6 \times 10^{-10}\ L_{44}^{1/3}$. However, when this happens the calculations above break down for another reason: CRs can freely stream from the source only if their Larmor radius in the pre-existing magnetic field is much larger than the assumed coherence scale of the field, namely
\be
E\gg 10^6 {\rm GeV}\ B_{-13} \lambda_{10}.
\label{eq:Larmor}
\ee
When Eq.~\ref{eq:Larmor} is not fulfilled, namely when the background field is relatively strong, then the propagation of CRs from the source becomes intrinsically one dimensional, which implies that the density of particles should be written as
\be
n_{\rm CR}(E,r) = \frac{Q(E) t}{\pi r_{L}^{2} v t} = \frac{2Q(E)}{\pi R_c(E)^2 c},
\ee
where we used the fact that the mean velocity of the particles is $v=c/2$ for a distribution of particles that is isotropic on a half plane and we assumed that particles spread in the direction perpendicular to the background field by a distance equal to $R_c(E)=\max(r_L(E), R_s)$ with $R_s$ the source size and $r_L(E)$ the Larmor radius of particles of given energy $E$. For a given source size $r_L>R_s$ as soon as $E_{\rm GeV} \gtrsim 9 \times 10^{6} B_{-10} (R_s/100 kpc)$. At energies larger than this:
\be
n_{\rm CR}(E,r)  \approx 47 L_{44}  E_{\rm GeV}^{-4} B_{-10}^{2} {\rm cm}^{-3} {\rm GeV}^{-1},
\ee
If particles with energy $>E$ are able to reach a given location, the current at that location is
\be
J_{\rm CR} \approx e \frac{c}{2} E n_{\rm CR}(E,r) = e  \frac{E Q(E)}{\pi r_{L}^{2}}\ ,
\ee
which results in non-resonant growth of the field for 
\be
E<E_{\rm inst}=3.5 \times 10^9 {\rm GeV}\ L_{44}^{1/2}\ ,
\label{eq:nonres2}
\ee
independent of $B_0$, and in a saturation magnetic field
\be
\delta B \approx 0.7\ E_{\rm GeV}^{-1}\ L_{44}^{1/2}\ B_{-10} ~ \rm Gauss.
\ee
This value of the magnetic field is apparently very large and reflects the very large density of particles at low energies in the proximity of the source, as due to the reduced dimensionality of the problem. However one should notice that the value is normalized to the density of $GeV$ particles, which only live in the immediate vicinity of the source and generate small scale fields to which high energy particles are almost insensitive. At Mpc scales, where only high energy particles can reach, the field strength is much lower as  we discuss below.

Assuming again that the diffusion coefficient is Bohm-like, one can write:
\be
D(E,r)=4.8\times 10^{16}\ E_{\rm GeV}^{2}\ L_{44}^{-1/2} B_{-10}^{-1} ~ \rm cm^{2}/s,
\ee
which leads to an estimate of the diffusion time:
$\tau_{\rm diff} = 6.2\times 10^{24}\ E_{\rm GeV}^{-2}\ L_{44}^{1/2}\ B_{-10}\ r_{\rm Mpc}^{2}~\rm yr.$
Requiring that particles reach the location at distance $r$ in a time shorter than the age of the universe, we then obtain:
\be
r_{\rm conf} \approx 0.5\  \left(\frac{E}{10^7\rm GeV}\right)\ L_{44}^{-1/4}\ B_{-10}^{-1/2}~ Mpc.
\label{eq:rconf2}
\ee
Following the usual procedure, one can calculate the growth rate of the fastest modes:
\be
\gamma_{\rm max} = \sqrt{\frac{4\pi}{\rho_{b}}}\frac{e c n_{\rm CR}(>E) }{2} = 1.9\times 10^{18} L_{44} B_{-10}^{2} E_{\rm GeV}^{-3} ~ \rm s^{-1},
\ee
and impose the condition that $\gamma_{\rm max}t_{0}>5$, which reads:
\be
E\lesssim E_{\rm growth} \approx 5.3\times 10^{11}\ {\rm GeV} L_{44}^{1/3} B_{-10}^{2/3}.
\label{eq:gmax2}
\ee

The intersection of all the conditions listed above leads to conclude that particles with energies 
\be
E<E_{\rm cut}=2.2\times 10^8 {\rm GeV}\ L_{44}^{1/4}\ B_{-10}^{1/2}\ \lambda_{10}
\ee
will be confined within a radius 
\be
r_{\rm conf}\approx 10\ {\rm Mpc}\ \lambda_{10}\ .
\ee
The amplified magnetic field at such distance is 
\be
\delta B\approx 3 \times 10^{-9} {\rm G}\ L_{44}^{1/4}\ B_{-10}^{1/2}\ \lambda_{10}^{-1}.
\ee
We emphasise again that the results illustrated both in the case of 3-d (lower $B_{0}$) or 1-d propagation (higher $B_{0}$) are only sensitive to the CR current, and hence insensitive to whether particle propagation is ballistic or diffusive. 

\section{Summary}

The strong effect that CRs have on the environment in which they propagate has long been investigated in the context of particle acceleration at shock waves \cite{bell78,bell2004} and is accompanied by observational consequences \cite{blasirev,amatorev}, such as spatially thin rims of enhanced X-ray synchrotron emission (see \cite{Vink} for a review). Here we investigated these processes in the immediate proximity of the sources (or host galaxies of the sources) of UHECRs, where the CR current generates instabilities that can change of way CRs propagate. 

For weak values of the strength of the pre-existing magnetic field $B_{0}$ (say $\lesssim 10^{-10}$ G), in the absence of non-linear phenomena, CRs propagate in approximately straight lines. The resulting electric current leads to the development of a Bell-like instability, that modifies the propagation of particles to be diffusive: we find that particles with energy $\lesssim 10^{7} L_{44}^{2/3}$ GeV are confined inside a distance of $\approx 3.8 L_{44}^{1/6}$ Mpc from the source for times exceeding the age of the Universe, thereby introducing a low-energy cutoff at such energy in the spectrum of CRs reaching the Earth. Since the confinement distance is weakly dependent on the source luminosity, we conclude that a region with $\sim nG$ fields should be present around any sufficiently powerful CR source. If larger background magnetic fields are present around the source, the gyration radius of the particles can be smaller than the coherence scale of the field, and in this case CR propagation develops in basically one spatial dimension. For a coherence scale of 10 Mpc, CRs are confined in the source proximity for energies $E\lesssim 2\times 10^{8} L_{44}^{1/4} B_{-10}^{1/2}\ \lambda_{10}$ GeV. 

Whether nature behaves in one or the other way depends on the poorly known value of $B_{0}$. Faraday rotation measures \cite{burles} provide weak, model dependent upper limits to $B_{0}$ in the nG range, while lower limits can be imposed based on gamma ray observations of distant TeV sources \cite{nero1,nero2} ($B_{0}\gtrsim 10^{-17}$ G). 

The physical prescription adopted here leads to estimating the strength of the self-generated magnetic field $\delta B$ in the source proximity at the level of equipartition with the energy in the form of escaping cosmic rays, independent of the value of the pre-existing field, $B_{0}$. A weak dependence on $B_{0}$ was instead found for the saturation level in \cite{riquelme}, which in our case would lead to $\delta B$ about $\sim 10$ times smaller for small values of $B_{0}$, thereby reducing the energy below which CRs are confined in the source proximity. Understanding the dynamics of the magnetic field amplification and saturation is clearly very important. One could test the amplification mechanism in the case of supernova remnant shocks: in this case the saturation criterion used here translates to $\delta B\approx \sqrt{4 \pi w_{\rm CR} v_S/c}$, with $v_S$ the velocity of the supernova blast wave and $w_{\rm CR}$ is the energy density in accelerated particles. Applying this criterion, we obtain an estimate of the magnetic field which is in good agreement with that measured in young galactic SNRs \cite{Vink}. On the other hand, due to the relatively small value of $\delta B/B_{0}$, the saturation provided by \cite{riquelme} would return a value of $\delta B$ only a factor $\sim 2$ smaller, too small a difference to discriminate between the two estimates. The testing is then left to numerical experiments studying the propagation of a current of energetic particles in a low density, low magnetic field plasma: hybrid simulations with this aim are currently ongoing.

The phenomenon of CR confinement illustrated here has profound implications for the description of the transition region between Galactic and extra-galactic CRs \cite{dip1,dip2,mixed}. It is rather tantalising that the cutoff obtained here as due to self-trapping is in the same range of values that have previously been invoked in the literature based upon phenomenological considerations.

\end{document}